\documentclass[final]{svjour3}
\usepackage{rotating}
\usepackage{amssymb}
\usepackage{mathptmx}
\usepackage{hyperref}
\hypersetup{
     colorlinks   = true,
     allcolors    =blue
}
\usepackage{graphicx}
\usepackage{caption}
\usepackage{tabularx}
\usepackage[numbers, square, sort]{natbib}
\makeatletter
\journalname{Journal of Low Temperature Physics}

\bibpunct{}{}{,}{s}{}{,}

\newcommand*{\citen}[1]{%
  \begingroup
    \romannumeral-`\x 
    \setcitestyle{numbers}%
    \setcitestyle{square}
    \hspace{-0.5em}\cite{#1}%
  \endgroup   
}

\usepackage{etoolbox}

\begin{document}

\AtEndEnvironment{thebibliography}{
\bibitem{simon/2017}S.M. Simon, et al., \textit{Journal of Low Temperature Physics}, in this Special Issue LTD17.
\bibitem{koopman/2017}B.J. Koopman, et al., \textit{Journal of Low Temperature Physics}, in this Special Issue LTD17.
}

\newcommand{\hdblarrow}{H\makebox[0.9ex][l]{$\downdownarrows$}-}
\title{Characterization of the Mid-Frequency Arrays for Advanced ACTPol}

\author{S.K. Choi \textsuperscript{1} \and 
J. Austermann \textsuperscript{2}  \and  
J.A. Beall \textsuperscript{2} \and 
K.T. Crowley \textsuperscript{1} \and 
R. Datta \textsuperscript{3, 4} \and 
S.M. Duff \textsuperscript{2} \and 
P.A. Gallardo \textsuperscript{5} \and 
S.P. Ho \textsuperscript{1} \and 
J. Hubmayr \textsuperscript{2} \and 
B.J. Koopman \textsuperscript{5} \and 
Y. Li \textsuperscript{1} \and
F. Nati \textsuperscript{6} \and
M.D. Niemack \textsuperscript{5} \and 
L.A. Page \textsuperscript{1} \and 
M. Salatino \textsuperscript{1} \and 
S.M. Simon \textsuperscript{3} \and 
S.T. Staggs \textsuperscript{1} \and 
J. Stevens \textsuperscript{5} \and 
J. Ullom \textsuperscript{2} \and 
E.J. Wollack \textsuperscript{4}}
 
\institute{\textsuperscript{1}Joseph Henry Laboratories of Physics, Jadwin Hall, Princeton University, Princeton, NJ 08544, USA \\
\textsuperscript{2}Quantum Sensors Group, NIST, Boulder, CO 80305, USA \\
\textsuperscript{3}Department of Physics, University of Michigan, Ann Arbor, MI 48103, USA \\
\textsuperscript{4}NASA Goddard Space Flight Center, Greenbelt, MD 20771, USA \\
\textsuperscript{5}Department of Physics, Cornell University, Ithaca, NY 14853, USA \\
\textsuperscript{6}Department of Physics, University of Pennsylvania, Philadelphia, PA 19104, USA  \\
\email{khc@princeton.edu}
}

\authorrunning
\titlerunning
\maketitle

\begin{abstract}

The Advanced ACTPol upgrade on the Atacama Cosmology Telescope aims to improve the measurement of the cosmic microwave background anisotropies and polarization, using four new dichroic detector arrays fabricated on 150-mm silicon wafers. These bolometric cameras use AlMn transition-edge sensors, coupled to feedhorns with orthomode transducers for polarization sensitivity. The first deployed camera is sensitive to both 150 GHz and 230 GHz. Here we present the lab characterization of the thermal parameters and optical efficiencies for the two newest fielded arrays, each sensitive to both 90 GHz and 150 GHz. We provide assessments of the parameter uniformity across each array with evaluation of systematic uncertainties. 
Lastly, we show the arrays' initial performance in the field.

\keywords{cosmic microwave background, transition-edge sensor, Atacama Cosmology Telescope, Advanced ACTPol, detector efficiency}
\vspace{-0.1in} 
\end{abstract}

\vspace{-0.15in}
\section{Introduction}
Low temperature detectors have enabled observations of the cosmic microwave background (CMB) radiation that has led to the most precise constraints yet for many of the cosmological parameters \citen{planck/2016b}. In recent years transition-edge sensor (TES) detector arrays have rapidly advanced the measurement of the CMB polarization. Increasing the number of detectors and frequency channels is a common theme for the CMB experiments today \citen{henderson/2016b, abazajian/2016} in order to more precisely characterize the faint CMB polarization buried among foreground emissions. 

The Atacama Cosmology Telescope (ACT), located at 5190 m altitude in Chile, measures the CMB anisotropies with a $\sim1.4'$ beam (at 150 GHz) \citen{thornton/2016}. The Advanced ACTPol (AdvACT) upgrade comprises four new polarization-sensitive dichroic detector arrays. They approximately double the number of detectors, each with improved sensitivity, and add three more frequency channels compared to the previous ACTPol receiver \citen{henderson/2016b}. These new bolometric detector arrays, fabricated at NIST, use AlMn \citen{dale_li/2016} as the superconducting metal and operate at $\sim 100$ mK. The first high-frequency (HF) array, sensitive to both 150 GHz and 230 GHz, has been deployed and operating since July 2016. Characterization and performance of the HF array are found in \citen{ho/2016}. Here we present the characterization and performance of two mid-frequency (MF) arrays fielded in April 2017, each sensitive to both 90 GHz and 150 GHz. In particular, we describe the lab characterization of their thermal parameters and optical efficiencies, and show the initial field performance. 

\vspace{-0.15in} 
\section{Instrument and Test Setup}
The four detector array cameras for AdvACT have largely common mechanical, electrical, and optical designs \citen{henderson/2016b, ward/2016}. Each array is fabricated on a 150-mm silicon wafer \citen{duff/2016} and consists of densely populated pixels coupled to light through a wide-band spline-profiled silicon feedhorn array \citen{simon/2016a}. Each detector element couples to the feedhorn with an orthomode transducer and has four AlMn TES bolometers for the two orthogonal polarizations and two frequency bands. Their frequency selectivity is defined with on-chip filters, a waveguide section on the feedhorn, and free-space metal-mesh filters. The arrays are read out with a two-stage Superconducting Quantum Interference Device (SQUID) system for time-division multiplexing \citen{irwin/2002, doriese/2016}. The arrays are integrated with the feedhorns, readout components, printed circuit board, mechanical supports, and flexible circuitry \citen{pappas/2016}, as described in \citen{li/2016, ward/2016}. 

Each MF array contains 1716 detectors in 429 pixels. The superconducting critical temperature, $T_c$, and thermal conductance to the bath, $G$, are tuned to optimize detector performance, accounting for the expected loading conditions during observations. 
We have also adjusted the heat capacity to tune the detector response time to avoid instability during operation. 

The first array (MF1) was fully integrated with the silicon feedhorn array prior to the in-lab characterization, whereas the second array (MF2) was first tested on a copper base, which was replaced with the silicon feedhorn array before deployment. The thermal bath environment is expected to be different for the two test setups given the heat capacity and conductivity differences between the silicon feedhorn and the copper base. We cover 2/3 of the pixels on the feedhorn array for MF1 and illuminate the other 1/3 with a cold load blackbody source to measure the optical efficiency. For MF2, 33 conical horns with $20^\circ$ flare angle are drilled into the copper base for cold load illumination. Each MF array is held fixed between the feedhorn array (or the copper base) and a metal clamp ring \citen{ward/2016}, to which we mount resistance thermometers (Ruthenium Oxide from LakeShore\footnote{http://www.lakeshore.com}) to measure the bath temperature for the detectors. 


%

\vspace{-0.15in}
\section{Data, Method, and Results}
\vspace{-0.15in}
\subsection{Thermal Parameters}
\label{sec:dark_param}
\begin{figure}
    \centering
    \includegraphics[width=1\textwidth]{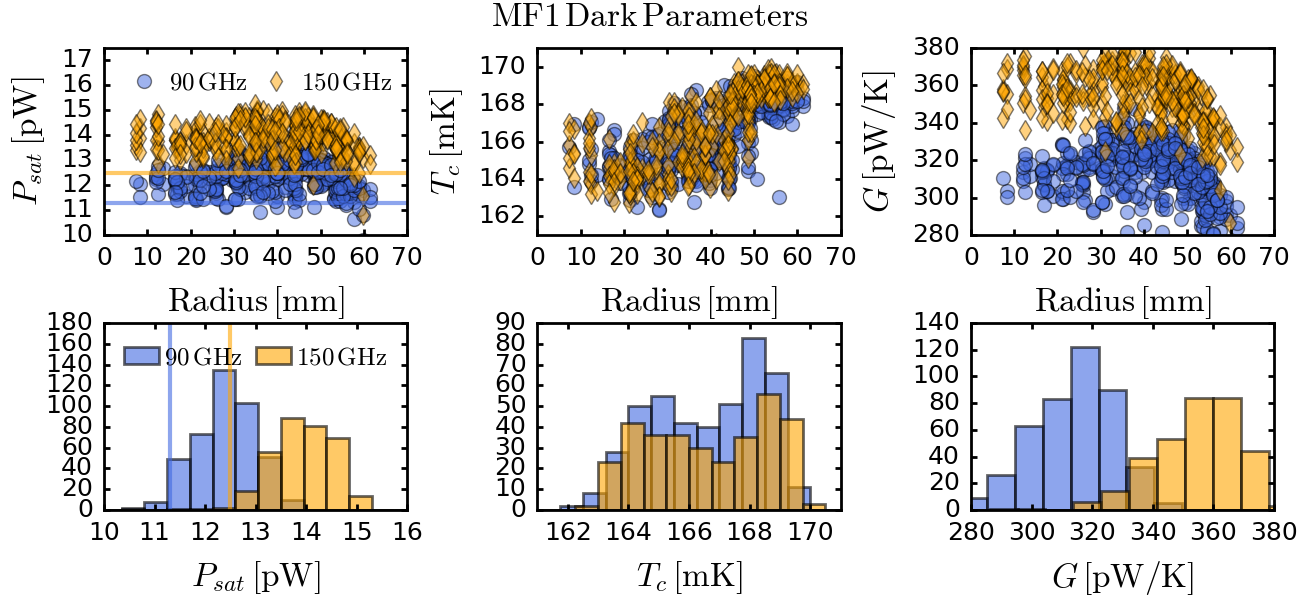}
    \\
    \vspace{0.cm}
    \includegraphics[width=1\textwidth]{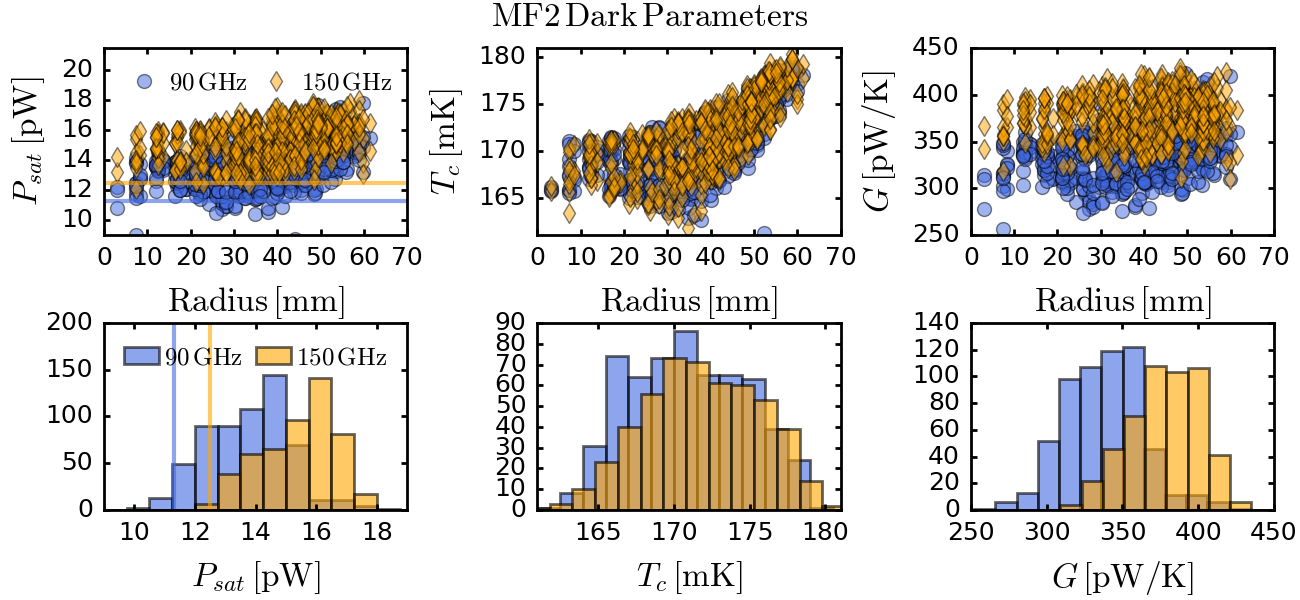}
\vspace{-0.25in} 
    \caption{Top panels show $P_{sat}$ at 100 mK (with targets shown in vertical/horizontal lines), $T_c$, and $G$ vs radius for the dark detectors, and the bottom panels show the corresponding histograms. The means and standard deviations are given in Table~\ref{tab:dark_param}. }
\label{fig:dark_param}
\vspace{-0.20in} 
\end{figure}

The data acquisition and analysis methods for detector thermal parameter characterization largely follow \citen{ho/2016}. In this section, we consider only the detectors covered from the cold load illumination, denoted ``dark." Raw data consist of $P_{sat}$, the electrical bias power at which the detectors are driven to 90\% of TES normal resistance, measured at various bath temperatures, $T_{bath}$. To measure $P_{sat}$ at each $T_{bath}$, we acquire current-voltage (IV) curves by first ramping the voltage bias to drive the detectors normal, then sweeping the bias down through the superconducting transition. We assume the usual power-law dependence for the power conducted to the bath from the TES, 
\vspace{-0.050in} 
\begin{equation}
P_{bath}(T) = K(T^n - T_{bath}^n),
\vspace{-0.050in} 
\label{eq:psat}
\end{equation}
where T is the temperature of the TES, $K$ is a constant, $n$ is the power law index, and $P_{bath}(T_c) \equiv P_{sat}$. The thermal conductance to the bath is defined at $T_c$,
\vspace{-0.050in} 
$$G \equiv dP_{bath}(T)/dT|_{T_c} = n K T_c^{n-1}.\vspace{-0.050in} $$
Depending on the bath temperature, the first bias ramp for the IV curve acquisition, which drives the detectors normal over the whole array, can lead to momentary heating of the bath. To mitigate this, we take IV curves on a quarter of the array at a time. This causes less than 2 mK fluctuation of the bath temperature for the entirety of the data acquisition. We measure $P_{sat}$ at 12 (11) $T_{bath}$ points between 55 (60) mK and 160 mK for MF1 (MF2). Assuming uniform uncertainties for the $P_{sat}$ values measured at different $T_{bath}$ points, we minimize the sum of the squared residuals for Equation~\ref{eq:psat} to yield the best-fit $T_c$, $K$, and $n$ for each detector. We use a SciPy minimization function (with Powell's method) that takes in initial guesses for the parameters to fit. Since $K$ and $n$ are related to the geometry and material for the legs, and $T_c$ is simply the critical temperature for the superconducting film, we do not anticipate correlations of $n$ or $K$ with $T_c$. The strong degeneracy between $n$ and $K$, and the form of Equation~\ref{eq:psat} make the fitting sensitive to the initial guesses. To illustrate the $K$-$n$ degeneracy, we fix $n$ (or $K$) to a single value within the range found here for all the dark detectors then fit again for $T_c$ and $K$ (or $n$). This 2-parameter fit yields $T_c$ and the derived $G$ that are within 2\% of the 3-parameter fit results. For HF, \citen{ho/2016} reported a difference in $n$ between the 150 GHz and the 230 GHz bands. Breaking the $K$-$n$ degeneracy requires measurement of $P_{sat}$ to $\sim1$\% uncertainty, and so this difference in $n$ was not significant. Here we report $G$ and $T_c$ rather than $K$, $n$, and $T_c$. Table~\ref{tab:dark_param} and Figure~\ref{fig:dark_param} show the measurement of $G$ and $T_c$ from the 3-parameter fit, and $P_{sat}$ at 100 mK (in the dark) with the design targets indicated. We also report the effective detector time constant $f_{3dB}$ measured from the response lag to a small square-wave excitation added to the DC TES bias. Parameters are largely uniform, and $P_{sat}$ values are near or above our design targets.

 \vspace{-0.1in} 
\begin{table}[b]
\begin{center}
\setlength{\tabcolsep}{5pt}
\begin{tabularx}{1\textwidth}{ c c c c c c c c c}
Array & $\nu$ & $T_c$ & $P_{sat}$ & $G$ & $f_{3dB}$ & $\eta_D$ & $\eta_T$ & NET \\
unit & (GHz) & (mK) & (pW) & (pW/K) & (Hz) & - & - & ($\mu\rm{K}\sqrt{s}$)\\
 \hline
 \hline
 MF1 & 90 & $167\pm2$ & 12.$4\pm0$.7 & $314\pm14$ & $122\pm28$ & $0.96\pm0.05$ & $0.26\pm0.06$ & 12 \\ 
 MF1 & 150 & $167\pm2$ & 13.$9\pm0$.6 & $354\pm15$ & $117\pm30$ & $0.71\pm0.03$ & $0.26\pm0.05$ & 14 \\ 
 
 MF2 & 90 & $171\pm4$ & 13.$7\pm1$.4 & $338\pm26$ & $118\pm30$ & $0.83\pm0.10$ & $0.30\pm0.07$ & 11 \\ 
 MF2 & 150 & $172\pm4$ & 15.$4\pm1$.2 & $378\pm23$ & $116\pm28$ & $0.66\pm0.04$ & $0.37\pm0.08$ & 12 \\ 
 \hline
\end{tabularx}
 \caption{The means and standard deviations of dark parameters and efficiencies are given for each frequency band for each array as described in Sections~\ref{sec:dark_param}, \ref{sec:opt_eff} and \ref{sec:field}. The right most column gives the array sensitivity at 1.3 mm of PWV divided by the sine of the observing altitude; see Section~\ref{sec:field}.}
 \label{tab:dark_param}
\end{center}
\end{table}

\vspace{-0.15in}
\subsection{Optical Efficiency}
\label{sec:opt_eff}
We estimate the optical efficiency for the detectors illuminated with the cold load by comparing the measured optical power to the expected power. The test setup and method are similar to \citen{crowley/2016}. The cold load source is a 15 cm by 15 cm wide aluminum slab with an array of square pyramids (height 2 cm and width 1 cm) cut on the surface and covered with Eccosorb CR-110. The source reflectivity was measured with a Vector Network Analyzer to infer the emissivity of $\sim1$ over the frequency bands of interests ($S21<-20$ dB). Three large diameter metal-mesh low-pass filters, with cutoffs at 5.9, 8.5 and 12 cm$^{-1}$, are mounted to reduce radiative loading on the cold stage. Due to space constraints given the sizes of the detector array, cold load, and the low-pass filters, the cold load is placed 13.5 cm away on-axis from the feedhorn, and hence does not fill the detector beams. Depending on the pixel position, the cold load subtends between 66 (53) and 92 (72)\% of the total solid angle for the 150 (90) GHz detectors for MF1 (and higher for MF2). The expected power depends on the cold load temperature $T_{CL}$, transmission of the filters, the cold load emissivity, and the solid angle subtended by the cold load at each pixel position $p$ given the simulated beam for the feedhorn. We compute the power at $p$ using the expression,
\vspace{-0.050in} 
$$P_p(T_{CL}) = \frac{1}{2}\int \epsilon(\nu)f(\nu)A_{e}(\nu)P(\theta,\phi,\nu)B(\nu,T_{{CL}})d\Omega d\nu, \vspace{-0.050in} $$ 
where 1/2 is for one polarization, $\epsilon(\nu)$ is the cold load emissivity, $f(\nu)$ is the product of the transmission curves for the three filters, $A_{e}(\nu)$ is the effective area for the feedhorn, $P(\theta,\phi,\nu)$ is the normalized beam pattern from HFSS\footnote{http://www.ansys.com/Products/Electronics/ANSYS-HFSS} simulations \citen{simon/2016a}, and $B(\nu, T_{{CL}})$ is the Planck function. As $A_e(\nu)$ is equal to $\lambda^2/\Omega$ for single-moded operation, we can integrate the angular parts and get:
\vspace{-0.050in} 
$$P_p(T_{{CL}}) = \frac{1}{2}\int \epsilon(\nu)f(\nu)\frac{c^2}{\nu^2}\frac{\Omega_{p}(\nu)}{\Omega(\nu)}B(\nu,T_{{CL}}) d\nu,\vspace{-0.050in} $$
where $\Omega_{p}(\nu)$ is the solid angle subtended by the cold load at $p$, and $\Omega(\nu)$ is the total solid angle of the feedhorn beam.

With the bath temperature held fixed at 100 mK, we step $T_{CL}$ at 9 points from $\sim9$ K to $\sim22$ K and measure $P_{sat}$ for all detectors. We servo to hold $T_{CL}$ constant at each step within $\sim4$ mK. The $P_{sat}$ of each illuminated detectors is lowered by the amount of optical power from the cold load. We measure this optical power by subtracting $P_{sat}$ of each illuminated detector from $P_{sat}$ of the dark detectors at the same radius within an annulus of width 1 cm, assuming azimuthal symmetry. Changing the annulus width by $\pm0.2$ cm corresponds to $<0.4\%$ difference in the final estimated efficiency. For MF1, where we had the detector array placed on the silicon feedhorn array, the $P_{sat}$ for the dark detectors decreases by $\sim1.2$ pW as $T_{CL}$ changes from 10.4 K to 22 K, although the temperature of the bath is held fixed at 100 mK within 2 mK. From Equation~\ref{eq:psat}, this corresponds to the cold load radiation causing a $\sim10$ mK gradient between the detector array and the point at which we measure its temperature (on the metal ring from which the silicon assembly is suspended). For MF2, where we had the detector array placed on a copper base, we do not see this gradient: $P_{sat}$ for the dark detectors stays constant within 0.1 pW as $T_{CL}$ changes from 8.5 K to 22 K. 

\begin{figure}
    \centering
    \includegraphics[width=1\textwidth]{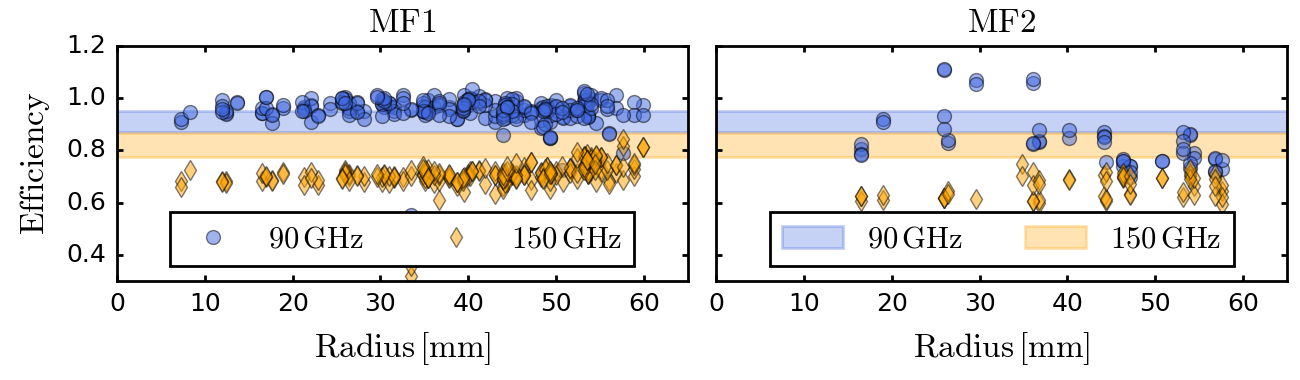}
\vspace{-0.25in} 
    \caption{Optical efficiency estimated from cold load measurements shown vs radius, with the 90 GHz detectors in blue circles and the 150 GHz detectors in yellow diamonds. The horizontal bands represent the expected efficiency from integrating the simulated passbands over all frequencies then dividing by the integral of top hats with the frequency bounds given in Section 3.2. The width of the band represents the expected range of loss in the silicon nitride lines with $0.0008 < \tan\delta < 0.002$. For the MF2 measurements, metal feedhorns were used that only illuminated a small number of the detectors in the array.}
\label{fig:opt_eff}
\vspace{-0.20in} 
\end{figure}

We define the optical efficiency relative to the calculated power for a top hat between 128.5 GHz and 168.5 GHz for the 150 GHz band and 78 GHz and 107 GHz for the 90 GHz band, chosen from the FWHM of the passbands simulated in Sonnet\footnote{http://www.sonnetsoftware.com} using transmission line modeling of the circuit. We estimate the detector optical efficiency $\eta_D$ from the slope of the measured and the calculated powers with,
\vspace{-0.050in} 
\begin{equation}
P^{\rm{meas}}_p(T_{\rm{CL}}) = \eta_D P_p(T_{\rm{CL}}) + C, 
\vspace{-0.050in} \vspace{-0.050in}
\label{eq:eff}
\end{equation}
where $P^{\rm{meas}}_p(T_{\rm{CL}})$ is the measured optical power at position $p$ for $T_{CL}$, and $C$ is a constant, which absorbs any power offset independent of $T_{CL}$. The estimated optical efficiencies are shown in Table~\ref{tab:dark_param} and Figure~\ref{fig:opt_eff}. The horizontal bands represent the expected efficiency calculated from integrating the Sonnet-simulated passbands over all frequencies then dividing by the integral of the appropriate top hat for each frequency band. The width of the band represents the expected range of loss in the silicon nitride lines assuming $0.0008 < \tan\delta < 0.002$ \citen{cataldo/2012}. We find that the mean of the 90 GHz efficiencies is $\sim5$\% higher and the mean of the 150 GHz efficiencies is $\sim15$\% lower than the center of the corresponding horizontal bands from simulations for each frequency. This may be an indication of slight differences in the actual passbands of the detectors compared to the Sonnet-simulated passbands. Data from an \textit{in situ} Fourier transform spectrometer (FTS) measurement is currently being analyzed.

For both MF1 and MF2, where the corrections in the dark $P_{sat}$ due to the changing bath temperature for increasing $T_{CL}$ differed, we find good linear fits for Equation~\ref{eq:eff}. Had the dark detectors behaved differently than the optical detectors, we would find a deviation between the two setups in the trend of $P_{sat}$ with $T_{CL}$. Other possible sources of systematic errors include not accounting for the solid angle outside the direct cold load emission, that can be filled by the emission of the shells or any reflected cold load radiation. For the former, we have 1 K emission through one low-pass filter and 4 K emission through three low-pass filters, totaling at most 0.35 pW for 90 GHz and less for 150 GHz. Note that this is independent of $T_{CL}$, and hence absorbed by the constant $C$ in Equation~\ref{eq:eff}. 
Any reflected cold load emission can in principle add to the measured optical power to potentially bias our measurement. Within the 13.5 cm distance between the cold load and the feedhorn, there are two $\sim1$ cm long cylindrical rings (filter clamps) at 4 cm and 9 cm away from the feedhorn, which could possibly reflect the cold load radiation to the feedhorns. When not accounting for the feedhorn beam, the sum of the solid angle of the two rings is $\sim15$\% of the solid angle subtended by the cold load at the center of the feedhorn. Including the feedhorn beam pattern will reduce this solid angle. Furthermore, there are only a few pixel positions that can receive the reflected cold load radiation directly given the geometry. We do not see strong evidence for such contamination: for MF1, the efficiencies are unchanging over radius. Thus we conservatively assume the reflected power can account for at most 15\% of the measured optical power, translating to at most 15\% overestimation of the efficiencies.

\vspace{-0.15in}
\subsection{Performance \textit{in situ}}
\label{sec:field}
After finding the dark and optical parameters within target, we deployed MF1 and MF2 to ACT, replacing the previous-generation detector arrays. 
The yields of working detectors on the telescope are $\sim88$\% for MF1 and $\sim76$\% for MF2. Most of the loss comes from cutting individuals channels with bad or no SQUID curves. For MF2, some of the loss is due to some persistent SQUIDs (that cannot be turned off) and a bias line short in the array.

\begin{figure}
    \centering
    \includegraphics[width=1\textwidth]{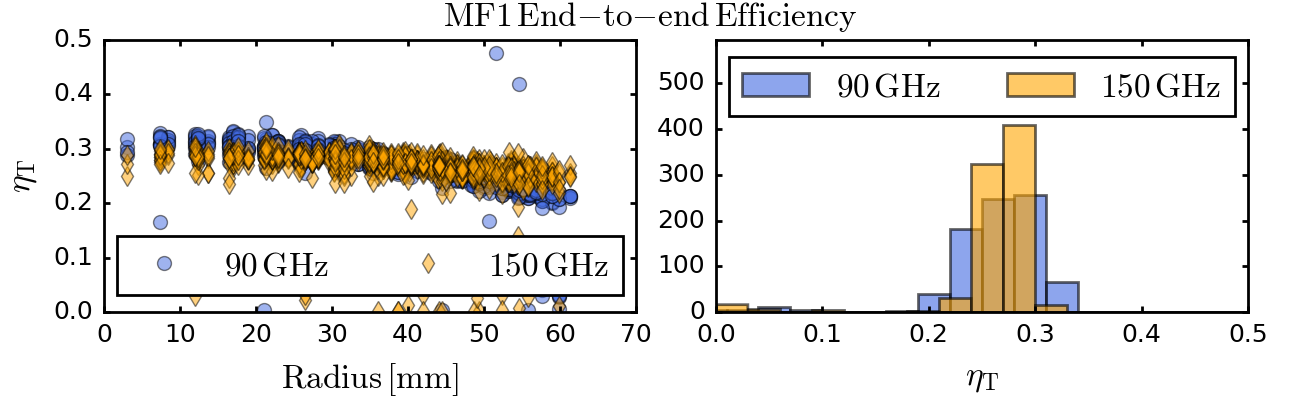}
    \\
    \vspace{0.2cm}
    \includegraphics[width=1\textwidth]{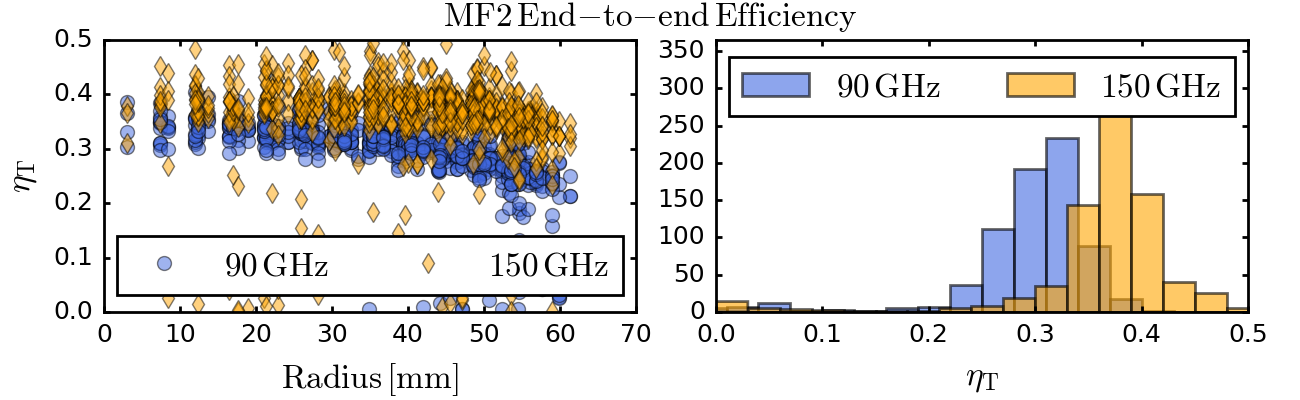}
    \vspace{-0.25in} 
    \caption{Telescope end-to-end efficiency estimated from Uranus observations using Equation~\ref{eq:eff_T}, with the 90 GHz detectors in blue circles and the 150 GHz detectors in yellow diamonds on the left panels. Corresponding histograms shown on the right panels. Clear decreasing trends are seen versus radius as $\eta_T$ includes light loss from truncating the feedhorn beams at the Lyot stop.}
\label{fig:eff_T}
\vspace{-0.20in} 
\end{figure}

We observe Uranus to calibrate the electrical response of the detectors to temperature fluctuations on the sky \citen{hasselfield/2013}. Uranus is effectively a point source for ACT, and we are sensitive to its diluted temperature $T_m$ given by $T_m = T_p \Omega_p/\Omega_B,$ where $T_p$ is the temperature of the planet, $\Omega_p$ is the apparent solid angle of the planet, and $\Omega_B$ is the solid angle of the instrument beam. For each detector, which is single-moded and measures one linear polarization, the expected optical power from such a source is given by, 
\begin{equation}
P_S = k_B T_m \Delta\nu \eta_T ,
\label{eq:eff_T}
\end{equation}
where $k_B$ is the Boltzmann constant, $\Delta\nu$ is the frequency bandwidth, and $\eta_T$ is the end-to-end efficiency of the telescope. For $\Delta\nu$, we use 26 (29) GHz for the MF1 (MF2) 90 GHz detectors and 36 (39) GHz for the respective 150 GHz detectors, based on \textit{in situ} FTS measurements. Then we measure $P_S$ from Uranus to estimate $\eta_T$ from each detector (Table~\ref{tab:dark_param}, Figure~\ref{fig:eff_T}). There is $\sim10$\% systematic uncertainty in the estimate of $\Omega_B$. As $\Omega_B$ is estimated from the Uranus map made with all detectors, $\eta_T$ also includes variations in the solid angle across the detectors, as well as signal losses from truncating the feedhorn beams at the Lyot stop, as seen in the downward slope with radius in the left panels of Figure~\ref{fig:eff_T}.

We also estimate the array sensitivity by measuring the detector noise power spectra then converting to antenna temperature units with the ratio $T_m/P_S$ from Uranus observations, followed by conversion to blackbody temperature units assuming frequency centers of 97 and 150 GHz. The white noise $n_i$ for each detector is estimated from a 4 Hz wide band centered at 20 Hz. The array sensitivity for each frequency band is obtained from 680--780 detectors with $(\sum_i n_i^{-2})^{-0.5}$. Array sensitivities, in noise equivalent temperature (NET), are estimated at 1.3 mm of precipitable water vapor (PWV) divided by the sine of the observing altitude, and reported in Table~\ref{tab:dark_param}. 
These sensitivity estimates are largely consistent with those inferred from noise levels measured in preliminary CMB maps.  

%
\vspace{-0.15in}
\section{Conclusion}
We have integrated, characterized, and deployed two new detector arrays for ACT. These mid-frequency arrays, sensitive to both 90 GHz and 150 GHz, are fabricated at NIST and shown to have uniform parameters near target. Initial field characterizations indicate good performance. Together with the high-frequency array (sensitive to both 150 GHz and 230 GHz) deployed last year, ACT is fully populated with the AdvACT detector arrays. The next array, sensitive to 27 GHz and 39 GHz \citen{simon/2017, koopman/2017}, is fabricated, assembled, and tested. It will be deployed in place of one of the MF arrays.

\vspace{-0.1in} 

\begin{acknowledgements}
This work was supported by the U.S. National Science Foundation through award 1440226. The development of multichroic detectors and lenses was supported by NASA grants NNX13AE56G and NNX14AB58G. The work of KTC and BJK was supported by NASA Space Technology Research Fellowship awards.
\end{acknowledgements}

\bibliographystyle{unsrt85}
\bibliography{publist}

\end{document}